\documentclass[11pt,aps]{article}
\usepackage{graphicx,epsfig,graphics,amssymb,amsmath,subeqnarray,color}

\def\deg{$^\circ$}
\def\d{{\rm d}}
\def\z{\zeta_w}
\def\z{\bar{z}}
\def\l{\lambda^*}

\def\e{\epsilon\epsilon_0}

\newcommand{\lsim}{\,\lower4pt\hbox{$\stackrel{{\textstyle <}}{\sim}$}\,}
\newcommand{\gsim}{\,\lower4pt\hbox{$\stackrel{{\textstyle >}}{\sim}$}\,}
\newcommand{\ggll}{\lower4pt\hbox{$\stackrel{{\gg}}{\ll}$}}

\begin{document}

\title{\textbf{ \Large Apparent slip due to the motion of suspended particles in flows of electrolyte solutions}}
\author{Eric Lauga\\
 {\it Division of Engineering and Applied Sciences, Harvard University},\\ {\it 29 Oxford Street, Cambridge, MA
 02138.}}
\date{\today}
\maketitle

\begin{abstract}
We consider  pressure-driven flows of electrolyte solutions  in
small channels or capillaries in which tracer particles are used
to probe velocity profiles. Under the assumption that the double
layer is thin compared to the channel dimensions, we show that the
flow-induced streaming electric field can create an apparent slip
velocity for the motion of the particles, even if the flow
velocity still satisfies the no-slip boundary condition. In this
case, tracking of particle would lead to the wrong conclusion that
the no-slip boundary condition is violated. We evaluate the
apparent slip length, compare with experiments, and discuss the
implications of these results.

\end{abstract}
\section{Introduction}

The no-slip boundary condition of fluid mechanics states that the
velocity of a viscous flow vanishes near a stationary solid
surface \cite{batchelor}. Although it has been a crucial
ingredient of our understanding of fluid mechanics for more than a
century, it has been much debated in the past \cite{Goldstein},
and, in the case of liquids, a complete physical picture for its
origin has yet to be given. The ongoing debate stems from the fact
that it is an assumption which cannot be derived from first
principles. It has been shown that on length scales much larger
than the scale of surface heterogeneities, the no-slip condition
might be a macroscopic consequence of inevitable microscopic
roughness \cite{Richardson,Jansons}, but the case of perfectly
smooth surfaces has yet to be explained. In particular, the
physico-chemical properties of both the fluid and the solid
surface certainly are important.

Only a few experimental studies have addressed the no-slip
condition in the past \cite{Schnell,Churaev}, and it is only the
recent advances in the controlled fabrication of micro- and
nano-devices and in the corresponding measurement techniques that
have allowed the problem to be reconsidered. Over the last few
years, a number of pressure-driven flow
 \cite{Watanabe,Cheng,Meinhart,Breuer}, shear-flow
\cite{Pit}, and squeeze-flow  experiments
\cite{Baudry,Craig,Bonaccurso,Cottin,Granick,Granick2002,Bonaccurso2}
showing a response interpretable as some degree of slip for
partially wetting liquids have been reported. Molecular dynamics
simulations of Lennard-Jones liquids have also shown that slip can
occur, but only at unrealistically high shear rates
\cite{Nature,Barrat}.

Fluid slip is usually quantified by a slip length $\lambda$. Let
us consider for simplicity a unidirectional flow past a solid
surface. Following Navier \cite{Navier}, the slip length linearly
relates the surface slip velocity to the shear rate of the fluid
evaluated at the surface
\begin{equation}\label{navier}
u=\lambda \frac{\partial u}{\partial n}\cdot
\end{equation}
The slip length can also be interpreted as the fictitious distance
below the surface at which the velocity would be equal to zero if
extrapolated linearly: the no-slip boundary condition is
equivalent to $\lambda=0$ and a no-shear boundary condition is
equivalent $\lambda=\infty$.

Consider pressure-driven flow in a two-dimensional channel of
height $2h$. If we assume that the boundary condition on the
channel walls ($z=\pm h$) is given by \eqref{navier}, the axial
velocity profile in the channel is
\begin{equation}\label{slip}
U_{\rm slip}(z)= -\frac{h^2}{2\mu}\frac{\d p}{\d
x}\left[1-\frac{z^2}{h^2}  +\frac{2\lambda}{h}\right],
\end{equation}
which is a Poiseuille flow augmented by a finite plug velocity,
which augmented flow rate ${Q_{\rm slip}}$ is given in a
non-dimensional form by
\begin{equation}\label{rate}
\frac{Q_{\rm slip}}{Q_{\rm no\mbox{\footnotesize
-}slip}}=1+\frac{3\lambda}{h}\cdot
\end{equation}

Experimentalists have usually addressed the issue of fluid slip in
two distinct ways. The first consists in performing indirect
measurements, such as pressure-drop versus flow rate or squeezing
rate versus resistance, and then use such measurements to infer a
slip length. This procedure is indirect in the sense that it
assumes that the flow resembles \eqref{slip} and then equation
\eqref{rate}, or an equivalent, is used to determine $\lambda$
\cite{Watanabe,Cheng,Breuer,Baudry,Craig,Bonaccurso,Cottin,Granick,Granick2002,Bonaccurso2}.

The second way consists in performing direct velocity measurements
in the fluid. We are only aware of two such previous works. Pit
{\it et al.} \cite{Pit} measured velocities in shear flow of
hexadecane over a smooth surface using a technique based on
fluorescence recovery after photobleaching (see also
\cite{Leger}). The measurements were performed down to 80 nm from
the solid surface and averaged over a few tens of microns. Fluid
slip was observed with $\lambda\sim 100$ nm in the case of
lyophobic surfaces. Tretheway \& Meinhart (2001) \cite{Meinhart}
used micro-particle image velocimetry (PIV) techniques to measure
the velocities of tracer nanoparticles (radius 150 nm) in
pressure-driven channel flow of water. Measurements were made down
to 450 nm from the solid surface and cross-correlated to increase
signal-to-noise ratios. Results consistent with the no-slip
condition were obtained in completely wetting conditions, but
 slip with $\lambda\sim1$ $\mu$m was obtained when the
channel walls were treated to be hydrophobic.

In this paper, we wish to draw attention to some of the possible
consequences of this latter type of particle-based measurements.
We address theoretically a prototypical pressure-driven flow
experiment in small channels in the case where small tracer
particles are used to probe the fluid velocity. We show that if
electrical effects for both the channel and the particles are
properly taken into account, it is possible for the particles to
behave  as if they were advected by a flow with a finite non-zero
slip length, even if the velocity profile in the fluid surrounding
the particle does not violate the no-slip condition.

In the following section we summarize some important background
electrostatics and hydrodynamics results, derive the formulae in
the case of two dimensional channels  and introduce the
electroviscous effect. In section \ref{suspended} we present a
physical picture for the effect we report, derive the expressions
for the apparent slip lengths and give the conditions for the
occurrence of such slip. Finally, in section \ref{discussion} we
discuss implications of these results along with estimates of
their order of magnitude under typical experimental conditions and
compare with experiments.

\section{Flow of an electrolyte solution}
\label{physical}

The physical picture for the effect we wish to introduce relies on
the following known facts.

\subsection{Surface charge and electrostatics}
\label{surfacecharge}

A solid surface in contact with an electrolyte solution will in
general acquire a net charge, due for example to the ionization of
surface groups, ion adsorption and/or dissolution. This surface
charge is a thermodynamic property of the solid-electrolyte pair
and the reader is referred to \cite{saville,israelachvili} for
detailed presentations of the phenomenon. The equilibrium surface
potential is called the zeta potential $\zeta$.

Such surface charges are screened by a diffusive cloud of
counter-ions in the solution. At equilibrium, the electrostatic
potential $\psi$ in the electrolyte satisfies the
Poisson-Boltzmann equation which quantifies the balance between
purely electrostatic interactions and diffusion \cite{saville},
\begin{equation}\label{PB}
\nabla^2 \psi =\frac{2en_0}{\e}\sinh
\left(\frac{e\psi}{k_BT}\right),
\end{equation}
where we consider here for simplification only the case of
monovalent 1:1 ions, e.g. Na$^+$ and Cl$^-$ or OH$^-$ and H$^+$.

A convenient approximation usually made to solve \eqref{PB} is the
Debye-H\"uckel approximation \cite{saville,rice,hunter,probstein}
of small field strength, $|e\psi| \ll k_BT$, in which case the
equation simplifies to the linearized Poisson-Boltzmann equation
\begin{equation}\label{debye}
\nabla^2 \psi =\kappa^2\psi, \quad \kappa^{-1}=\left(\frac{\e
k_BT}{2e^2n_0}\right)^{1/2},
\end{equation}
where $\kappa^{-1}$ is the Debye screening length: it is the
typical length scale in the solution over which counter-ions
screen the charged solid surface, and beyond which the net charge
density is essentially zero.

However, \eqref{debye} is restricted to low surface potentials,
typically 20mV, which is a severe approximation. Let us consider
for simplicity the case of a two-dimensional channel of height
$2h$ in the $z$-direction and let us instead derive the solution
to \eqref{PB} for any value of the zeta potential at the wall
$\zeta_w$ but in the limit where the channel dimensions are much
larger than the double layers $\kappa h\gg 1$. This limit is
appropriate for channel sizes down to $h\approx 5$ $\mu$m in the
case of pure water, or even $h\approx 50$ nm in the case of tap
water.

Let us define the dimensionless potential $\phi=e\psi/k_BT$ and
the dimensionless vertical coordinate $\bar{z}=z/h$. In this case,
\eqref{PB} becomes
\begin{equation}\label{newPB}
\frac{1}{(\kappa h )^2} \frac{\d^2 \phi}{ \d\z^2} =\sinh \phi,
\end{equation}
with the boundary conditions $\phi(\z=\pm
1)=\phi_w=e\zeta_w/k_BT$.

Since $1/\kappa h\ll 1$, the solution to equation \eqref{newPB}
involves boundary layers near $z=\pm 1$. The outer solution
$\phi_{\rm \,out}$ is found by taking the limit $1/\kappa h=0$ in
\eqref{newPB} and we find $\phi_{\rm \,out}=0$. The inner solution
$\phi_{\rm\, in}$ is valid near the boundaries for $\kappa
h(1-|\z|)={\cal O}(1)$, in which case \eqref{newPB} reduces to the
Poisson-Boltzmann equation near an infinite plane, whose solution
is \cite{hunter}
\begin{equation}\label{inner}
\tanh\left(\frac{\phi_{\rm\, in}(\z)}{4} \right) = \tanh
\left(\frac{\phi_w}{4} \right) e^{-\kappa h (1-|\z|)}.
\end{equation}
Finally, since $\phi_{\rm \,out}=0$, the inner solution
\eqref{inner} is also equal to the composite solution $\phi(\z)$,
uniformly valid throughout the channel as $\kappa h \to \infty$,
at leading order in $1/\kappa h$. For convenience, equation
\eqref{inner} can be rewritten as
\begin{equation}\label{comp}
\phi(\z)=2\ln \left(\frac{1+t_w e^{-\kappa h (1-|\z|)}}{1-t_w
e^{-\kappa h (1-|\z|)}} \right),
\end{equation}
where we have defined $t_w=\tanh (e\zeta_w/4k_B T)$.

\subsection{Hydrodynamics and electrokinetics}
When a pressure-driven flow occurs in the channel, the fluid
velocity is unidirectional ${\bf U}=U(z){\bf e}_x$, where ${\bf
e}_x$ is the streamwise direction. In the absence of electrical
effects, the fluid velocity is simply Poiseuille's pressure-driven
formula \cite{batchelor}, which we will denote $U_{{\rm PD}}$, and
is given by
\begin{equation}\label{PD}
  U_{\rm PD} (z)=-\frac{h^2}{2\mu}\frac{\d p}{\d x}\left[1-\frac{z^2}{h^2}
  \right]\cdot
\end{equation}

Furthermore, if an external, or induced, electric field ${\bf
E}_{S}=E_{S}{\bf e}_x$ is also applied to the channel, the
presence of a net charge density near the solid surface moving in
response to the field leads to an additional velocity component
known as electroosmotic flow (EOF) \cite{saville}. It is directed
in the $x$-direction, is given by
\begin{equation}\label{EOF}
  U_{{\rm EOF}}(z)=\frac{\e E_{S}}{\mu}\Big[\psi(z)-\zeta_w
  \Big],
\end{equation}
and is valid for any value of $\zeta_w$.

\subsection{Streaming potential and electroviscous effect}
\label{stream}

As the electrolyte solution flows down a pressure gradient, the
cloud of counter-ions is advected by the flow and a streaming
current is established. If no short-circuit is present between the
two ends of the capillary, accumulation of charge sets up a
potential difference along the channel, termed the ``streaming
potential''. Such potential, or equivalently electric field,
opposes the mechanical transfer of charge by creating a reverse
conduction current through the bulk solution such that the total
net electric current is zero. This induced axial electric field
scales with the applied pressure gradient and leads to the
creation of an induced electroosmotic back-flow which effectively
slows down the fluid motion in the capillary: a smaller flow rate
for a given pressure drop is obtained than in the regular
Poiseuille case, as if the liquid had a higher shear viscosity
than expected. Consequently this effect is usually referred to as
the primary ``electroviscous effect''
\cite{burgeen,rice,levine,hunter,probstein}.

Let us consider the pressure-driven flow in a channel of height
$2h$ and width $w\gg h$ of the electrolyte solution with
electrostatic potential given by equation \eqref{inner}. We
calculate below the value of the steady-state streaming electric
field $E_S {\bf e}_x$ induced by the flow.

\paragraph{Pressure-driven current}

First, the pressure-driven motion of the screening cloud of
counter-ions near the charged surface leads to an
advection-of-charge electric current $I_{S}^{PD}$ given by
\begin{equation}\label{ISPDbl}
I_{S}^{\rm PD}=\int_{-h}^hw\rho_e(z)U_{{\rm PD}}(z)\d z=\frac{2\e
wh k_B T}{\mu e} \left(\frac{\d p}{\d x} \right)I_1,
\end{equation}
where we have used the electrostatic equation to relate the net
charge density in the liquid to the electrostatic potential,
${\rho_e}= -\e \nabla^2 \psi$ and where $I_1$ is given by
\begin{equation}\label{I1}
I_1=\phi_w-\int_0^1\phi(\z)\d \z,
\end{equation}
with the same dimensionless notations as in section
\ref{surfacecharge}. In the limit where $\kappa h \gg 1$, plugging
in the solution \eqref{inner} into \eqref{I1} leads to
\begin{equation}\label{I1limit}
I_1=\phi_w - \frac{2}{\kappa h} \int_0^{\kappa h}\ln
\left(\frac{1+t_we^{-x} }{1-t_we^{-x} }\right)\d x,
\end{equation}
so that
\begin{equation}\label{ISPD}
I_{S}^{\rm PD}=\frac{2\e wh\zeta_w}{\mu }\left(\frac{\d p}{\d
x}\right)\left[1+{\cal O}\left(\frac{1}{\kappa h}\right)\right].
\end{equation}

\paragraph{Electroosmotic current}

If an electric field is induced by the flow, the streaming current
has a second component $I_S^{{\rm EOF}}$, given by the advection
of counter-ions by the induced electroosmotic flow
\begin{equation}\label{ISEOFbl}
I_S^{{\rm EOF}}=\int_{-h}^hw\rho_e(z)U_{\rm EOF}(z)\d z
=\frac{2wE_S}{h\mu}\left(\frac{\e k_BT}{e}\right)^2I_2,
\end{equation}
where $I_2$ is given by
\begin{equation}\label{I2}
I_2=\int_0^1\left(\frac{\d\phi}{\d\z}\right)^2\d\z.
\end{equation}
In the limit where $\kappa h \gg 1$, the boundary layer solution
\eqref{inner} leads to the leading order expression for $I_2$ in
powers of $1/\kappa h$,
\begin{equation}\label{I2limit}
I_2=\frac{8\kappa h t_w^2 (1-e^{-2\kappa
h})}{(1-t_w^2)(1-t_w^2e^{-2\kappa h})},
\end{equation}
so that
\begin{equation}\label{ISEOF}
I_S^{{\rm EOF}}=\frac{16 w \kappa E_S }{\mu}\left(\frac{\e k_B
T}{e}\right)^2 \left(\frac{t_w^2}{1-t_w^2}\right) \left[1+ {\cal
O}\left(\frac{1}{\kappa h}\right)\right]\cdot
\end{equation}

\paragraph{Conduction current} Finally, in response to the electric field, a
conduction current $I_C$ is set up in the bulk of the solution; if
we denote by $\sigma$ the ionic conductivity of the electrolyte
(assumed to be constant), the conduction current is given by
\begin{equation}\label{}
  I_C=2hw\sigma E_S.
\end{equation}

\paragraph{Induced electric field}

If we investigate the steady-state motion of the electrolyte
solution, we require that there be no net electric current
\begin{equation}\label{}
  I_{S}^{\rm PD}+I_S^{{\rm EOF}}+I_{C}=0,
\end{equation}
which leads to the formula for the flow-induced streaming electric
field
\begin{equation}\label{ES}
E_s=-\frac{\d p}{\d
  x}\left(\frac{\e\zeta_w}{\sigma\mu}\right)\left[1+\frac{8\kappa}{\sigma\mu
  h}\left(\frac{\e k_B
T}{e}\right)^2 \left(\frac{t_w^2}{1-t_w^2}\right)\right]^{-1}
+{\cal O}\left(\frac{1}{\kappa h}\right)\cdot
\end{equation}
As expected, the induced field $E_S$ is proportional to the
applied pressure gradient\footnote{The effect of the streaming
electric field on the properties of the flow (the
``electroviscous'' effect) can be understood by evaluating the
total flow rate from both \eqref{PD} and \eqref{EOF} and, with
\eqref{ES}, rewriting it under the form of an effective Poiseuille
flow rate with a different effective shear viscosity $\mu_{\rm
eff}$ \cite{hunter}. We find that $\mu < \mu_{\rm eff}$ so that,
from the standpoint of flow rate versus pressure drop, the
electrical effect effectively increases the bulk viscosity of the
solution.}.

Note that within the Debye-H\"uckel approximation \eqref{debye},
the induced electric field can be calculated exactly for all
values of $\kappa h$ \cite{saville,rice,hunter,probstein} and  we
find
\begin{equation}\label{ESdebye}
  E_S=\frac{\d p}{\d x}
  \left(  \frac{\tanh\kappa h}{\kappa h} -1\right)
    \left[
  \frac{\sigma\mu}{\e \zeta_w}+\frac{\e\zeta_w\kappa}{4h}\left(\frac{\sinh 2\kappa h-2\kappa h}{(\cosh\kappa h)^2}
  \right)\right]^{-1}\cdot
\end{equation}
In the limits where $e|\zeta_w|/k_BT\ll 1$ ({\it i.e.} $t_w\ll 1$)
and $\kappa h \gg 1$, the expressions given by \eqref{ES} and
\eqref{ESdebye} agree and are given by
\begin{equation}\label{}
  E_S= -\frac{\d p}{\d
  x}\left(\frac{\e\zeta_w}{\sigma\mu}\right)\left[1+\frac{(\e\zeta_w)^2\kappa}{2\sigma\mu
  h}\right]^{-1}\cdot
\end{equation}

\section{Velocity of a suspended particle and apparent slip}
\label{suspended}
\subsection{Physical picture}

We now consider an experiment in which the above electric effects
are present. We elect to use small tracer particles to probe the
velocity profile, including possible fluid slip, as illustrated in
Figure \ref{figure}. For the same reason as for the capillary
surfaces, these particles will usually be charged in solution. As
they are advected by the fluid motion, they will also feel the
influence of the induced streaming electric field: consequently
their velocity will not only reproduce that of the fluid but will
also include an induced electrophoretic component \cite{saville},
proportional to their zeta potential and the streaming electric
field. If the zeta potential of a particle has a sign opposite to
that of the capillary surface, the particle will be slowed down by
the electric field. On the contrary, if the particle possesses a
potential of the same sign as the capillary surface, its
electrophoretic component will be in the streamwise direction;
furthermore, if its zeta potential is large enough, the
electrophoretic velocity of the particle will be able to overcome
the induced electroosmotic back-flow.

It then follows that there is a significant potential implication
of the induced electric field: if one were to conduct an
experiment in such conditions without considering any important
electrical effects, these particles would go faster than the
expected Poiseuille pressure-driven profile, leading to the
incorrect conclusion that the velocity profile has a non-zero slip
velocity at the wall. Thus, even if the flow satisfies the no-slip
condition, measurements of particle velocities would lead to
non-zero apparent slip lengths. We shall quantify this mechanism
in the following sections.

\subsection{Particle velocity}

\begin{figure}[t]
\centering
\includegraphics[width=.7\textwidth]{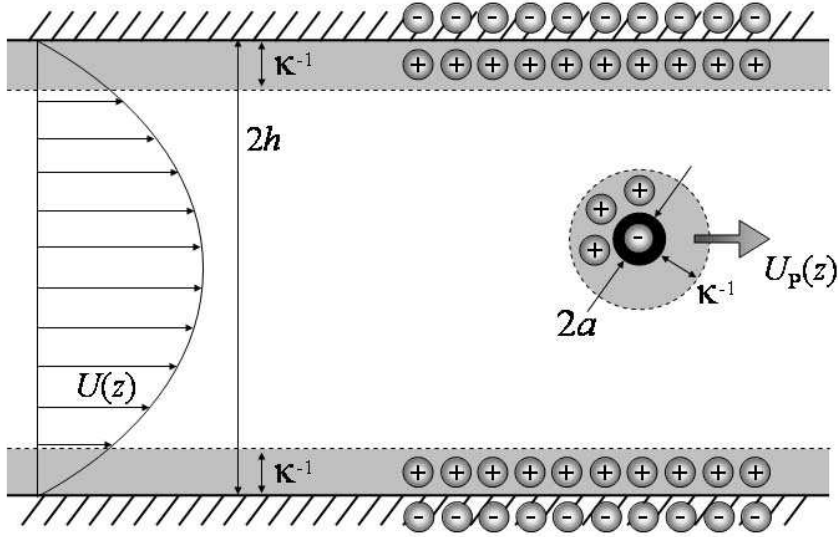}
\caption{Schematic representation of the flow between two parallel
plates with charged surfaces (zeta potential $\zeta_w$) and a
charged suspended particle (zeta potential $\zeta_p$); in the case
illustrated, $\zeta_w<0$ and $\zeta_p<0$. The channel height is
$2h$, the particle radius is $a$, the smallest wall-particle
distance is $d$ and the screening length $\kappa^{-1}$.}
\label{figure}
\end{figure}

We consider the presence of a single solid spherical particle of
radius $a\ll h$ suspended in a two-dimensional channel of height
$2h$ where a pressure-driven flow occurs, as illustrated in Figure
\ref{figure}; the particle is located at a distance $d=h-|z|$ from
the closest wall. We also assume for simplicity that the presence
of the particle does not modify the nature of ionic groups in
solution (1:1 monovalent ions), so that the screening lengths
$\kappa^{-1}$ for the charged particle and the charged channel
surface are the same, as given by equation \eqref{debye}.

The particle velocity ${\bf U}_{\rm P}(z)$ will in general be
\begin{equation}\label{Up}
{\bf U}_{\rm P}(z)= {\bf U}_{\rm hydro}(z)+{\bf U}_{\rm
elec}(z)+{\bf U}_{\rm k_B T},
\end{equation}
which includes three contributions.

\paragraph {Hydrodynamic contribution} The first component is the
hydrodynamic contribution
\begin{equation}\label{hydro}
{\bf U}_{\rm hydro}(z)=\left[1-{\cal O}\left(\frac{a}{d}\right)
\right] U_{\rm PD}(z){\bf e}_x,
\end{equation}
where $U_{\rm PD}$ is the local pressure-driven fluid velocity. It
is modified by the presence of solid walls which slow down the
motion of the suspended particle. Although the analysis is in
general difficult \cite{Happel}, walls lead to a leading-order
correction to the particle velocity of order of the ratio of the
particle size to the distance to the walls ${\cal O} (a/d)$; this
is true as long as the particle does not come too close to the
wall, in which case a different contribution arises from
lubrication forces. We will assume in this paper that the particle
is located sufficiently far from the walls ($a\ll d=h-|z|$) so
that the influence of the walls can be neglected. Such a
requirement would also have to be verified in an experiment,
otherwise the presence of the wall would hinder some component of
the measured slip velocity.  Note that if walls were not present,
a correction to the velocity accounting for the finite size of the
particle and the spatial variations of the fluid velocity would
also be present, but only at second order in the ratio of the
particle size to the length scale over which flow variations occur
\cite{hinch}.

\paragraph{Electrical contribution}
In general the particle will be charged, with a zeta potential
$\zeta_p$ which we assume to be uniform. Consequently, its
velocity will include a contribution from electrical forces, ${\bf
U}_{\rm elec}(z)$. This velocity has two components
\begin{equation}\label{}
  {\bf U}_{\rm elec}(z)={\bf U_{\rm EPH}}+U_{\rm drift}(z)\,{\bf
  e}_z,
\end{equation}
where ${\bf U_{\rm EPH}}$ is an electrophoretic velocity due to
the presence of an external electric field and $U_{\rm drift}(z)$
is a vertical drift due to the electrostatic interactions between
the charged particle and the charged walls. Such drift will only
be significant if the double layers around the particle and along
the channel walls overlap, and will be exponentially screened
otherwise \cite{saville}. We will assume that such requirement is
met in practice $\kappa d \gtrsim {\cal O}(1)$, so that it can be
neglected.

When the electric field ${\bf E}_S=E_S{\bf e}_x$ is aligned with
the channel direction, the electrophoretic velocity ${\bf U}_{\rm
EPH}=U_{\rm EPH}\,{\bf e}_x$ is given by
\begin{equation}\label{phoretic}
U_{\rm EPH}=\frac{\e E_S (f(\kappa a)\zeta_p-\zeta_w)}
{\mu}\left[1-{\cal O}\left(\frac{a^3}{d^3}\right) \right]\cdot
\end{equation}
This velocity first includes the ``pure'' electrophoretic mobility
of the particle \cite{saville,hunter,AnnRev}, characterized by the
function $f(x)$, which satisfies $f(0)=2/3$ (H\"uckel's result for
thick screening length) and $f(\infty)=1$ (Smoluchowski's result
for thin screening length). Note that we can use these classical
electrophoretic formulae because since $\kappa h \gg 1$, the
perturbation of the ion distribution in the double layer around
the particle is not modified by the local shear flow. The velocity
\eqref{phoretic} also includes the electroosmotic back-flow
resulting from the motion of excess charges near the channel walls
and proportional to the wall zeta potential $\zeta_w$.
Furthermore, the presence of a wall always influences the
electrophoretic mobility at cubic order in the ratio of the
particle size to the distance to the wall, as long as double
layers do not overlap \cite{Ennis,Yariv}; since we already assumed
the particle to be located far from the wall, we will neglect the
wall influence here as well.

\paragraph{Thermal contribution}
Finally, the particle velocity has a random contribution ${\bf
U}_{\rm k_B T}$ due to thermal motion, which can be significant. A
solid spherical particle of radius $a$, located far from
boundaries, has a diffusivity $D$ given by the Stokes-Einstein
relation $D=k_B T/ 6\pi \mu a$ \cite{saville}, corresponding to a
root mean square velocity on the order of $U_{k_BT} \sim D/a \sim
k_B T/ 6\pi \mu a^2$. At 25\deg C in water, $a= 10$ nm leads to
$U_{k_BT}\sim 1$ mm/s; this value is of the same order as the
fluid velocity in a circular capillary of radius $R\sim 100$
$\mu$m and flow rate $Q\sim 1$ $\mu$L/min, typical values for
microfluidic devices. Consequently, we cannot assume that the
Peclet number, $Pe=U/U_{k_B T}=Ua/D$, is necessarily large and
thermal motion cannot in general be neglected. However, in the
experiments reported to date, velocity measurements are cross
correlated (as in \cite{Meinhart}) or averaged (as in \cite{Pit})
so that the random thermal motion disappears, and we will
therefore not consider it in this paper.

\paragraph{Summary}
Under the previous assumptions, we can write the velocity for the
particle as
\begin{equation}\label{summary}
U_{\rm P}(z)  =U_{{\rm PD}}(z)+
 \frac{\e
E_{S}}{\mu}(f(\kappa a)\zeta_p-\zeta_w) + {\cal
O}\left(\frac{a}{d}\right),
\end{equation}
where the velocity should be understood as an ensemble average
over different experimental realizations.

\subsection{Apparent slip length}

We now calculate the apparent slip length $\lambda$ that would be
inferred by tracking particle motion in a pressure-driven flow. In
the limit $\kappa h \gg 1$, the streaming electric field is given
by equation \eqref{ES} so that the particle velocity
\eqref{summary} becomes, at leading order in $a/d$ and $1/\kappa
h$,
\begin{equation}\label{large1}
U_{\rm P}(z)  = -\frac{h^2}{2\mu}\frac{\d p}{\d
x}\left\{1-\frac{z^2}{h^2} +\frac{2\zeta_w(f(\kappa a
)\zeta_p-\zeta_w)(\e)^2}{\sigma\mu h^2}
\left[1+\frac{8\kappa}{\sigma\mu
  h}\left(\frac{\e k_B
T}{e}\right)^2
\left(\frac{t_w^2}{1-t_w^2}\right)\right]^{-1}\right\} \cdot
\end{equation}
Comparing \eqref{large1} with the formula for the velocity in a
flow satisfying the partial slip boundary condition \eqref{slip},
we see that the particle behaves as if it was passively advected
by a pressure-driven flow with a finite slip length $\lambda$
given by
\begin{equation}\label{slip1}
\frac{\lambda}{h}=\frac{\zeta_w(f(\kappa a)\zeta_p-\zeta_w)( \e
e)^2}{\sigma\mu (eh)^2+{8\kappa h}\left({\e k_B T}\right)^2
\left(\frac{t_w^2}{1-t_w^2}\right)}\cdot
\end{equation}
The condition for a positive apparent slip, $\lambda> 0$, is
therefore
\begin{equation}\label{condlarge1}
  \zeta_w(f(\kappa a)\zeta_p- \zeta_w)>0.
\end{equation}
This result can also be understood in the following way: (1) the
particle and the wall must have the same charge sign,
$\zeta_w\zeta_p>0$; this is usually the case in water where
surfaces typically acquire negative charge, for example due to the
ionization of sulfate or carboxylic surface groups; (2) the
particle zeta potential must be sufficiently large
$|\zeta_p|>|\zeta_w|/f(\kappa a)$ (or, equivalently, the wall zeta
potential must be sufficiently small). If condition
\eqref{condlarge1} is not met, the slip length is in fact a
``stick'' length ($\lambda<0$) and the particle goes slower than
the liquid. Finally, note that within the Debye-H\"uckel limit
$t_w\ll 1$, the slip length \eqref{slip1} becomes
\begin{equation}\label{slipdebye}
\frac{\lambda}{h}=\frac{2\zeta_w(f(\kappa
a)\zeta_p-\zeta_w)(\e)^2}{2\sigma\mu h^2+(\e\zeta_w)^2\kappa h}
\cdot
\end{equation}

\section{Discussion}
\label{discussion}

The results presented in the previous section allow one to
calculate, for a given set of experimentally determined material
and fluid parameters, the amount of apparent slip in the particle
velocity which is due to the streaming potential. We present in
this section some general observations on formula \eqref{slip1} as
well as an estimate for the order of magnitude of the effect in
water and a comparison with available experimental slip
measurements.

\subsection{Variations of the slip length}

All the variables in \eqref{slip1} can be made to vary
independently except for the screening length $\kappa^{-1}$ and
the bulk conductivity $\sigma$ which both depend on the ionic
strength of the solution. A simple estimate for the bulk
conductivity of a 1:1 solution is $ \sigma= {2 b n_0e^2}$ (see
e.g. \cite{probstein}), where $n_0$ is the bulk ion concentration
and $b$ is the ion mobility, which we approximate by the mobility
of a spherical particle, $b^{-1} \approx 6\pi \mu \ell$ where
$\ell$ is the effective ion size. Using equation \eqref{debye}, we
see that the conductivity and the screening length are related by
\begin{equation}\label{cond2}
\sigma \approx \frac{\e k_B T}{6 \pi \mu \ell} \kappa^2.
\end{equation}

Furthermore, since the conductivity $\sigma$ and the viscosity
$\mu$ only appear in \eqref{slip1} as their product, the estimate
\eqref{cond2} shows that the apparent slip length \eqref{slip1} is
in fact independent of the fluid viscosity. Moreover, since
$\kappa\sim n_0^{1/2}$ and $\sigma\sim n_0$, and since $f(\kappa
a)$ varies only weakly with $\kappa$, we see from \eqref{slip1}
that the $\lambda$ is a decreasing function of the ionic strength.
Also, it is clear from \eqref{slip1} that the slip length always
decreases with the channel size.

\begin{figure}[t] \centering
\includegraphics[width=.48\textwidth]{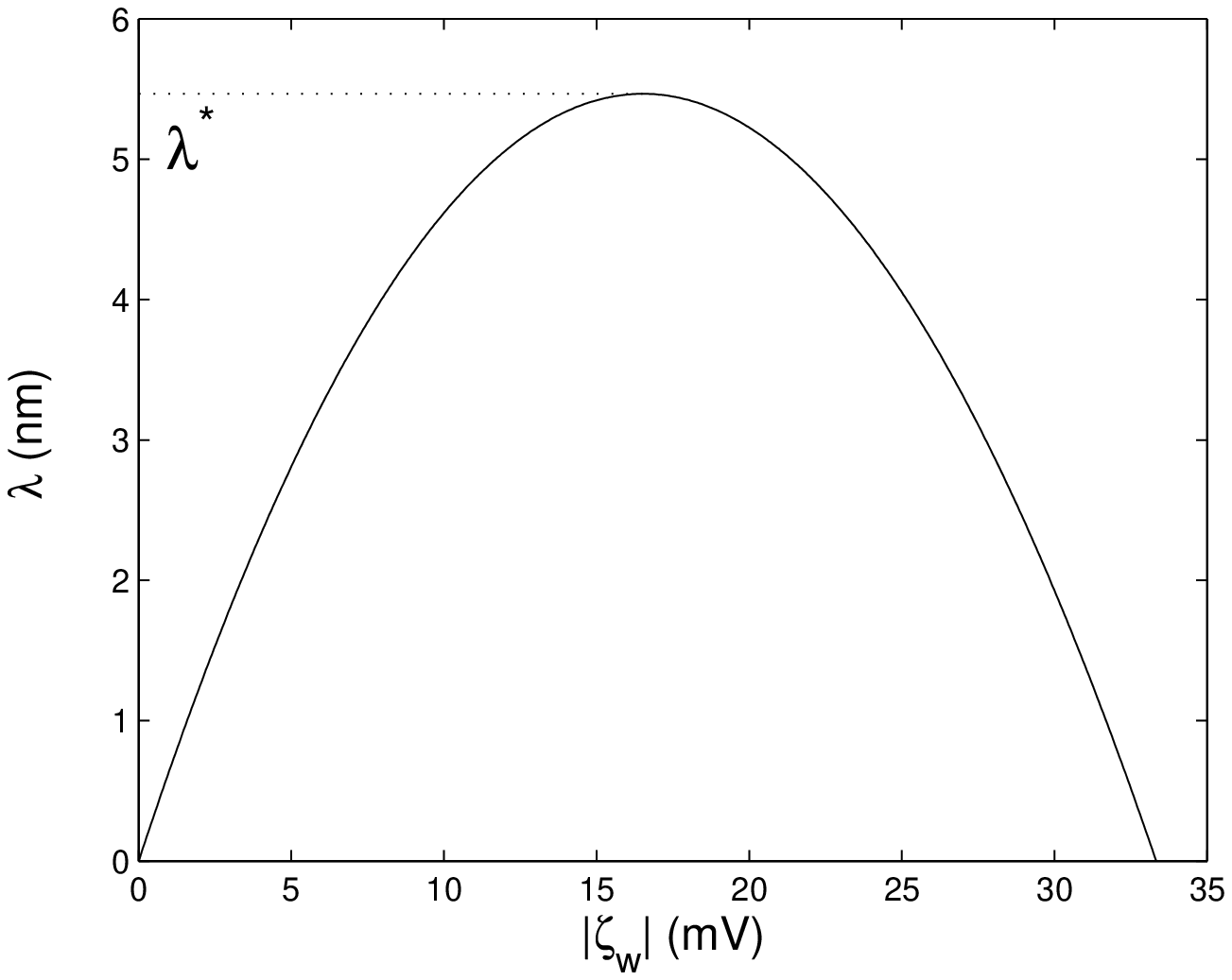}
\includegraphics[width=.48\textwidth]{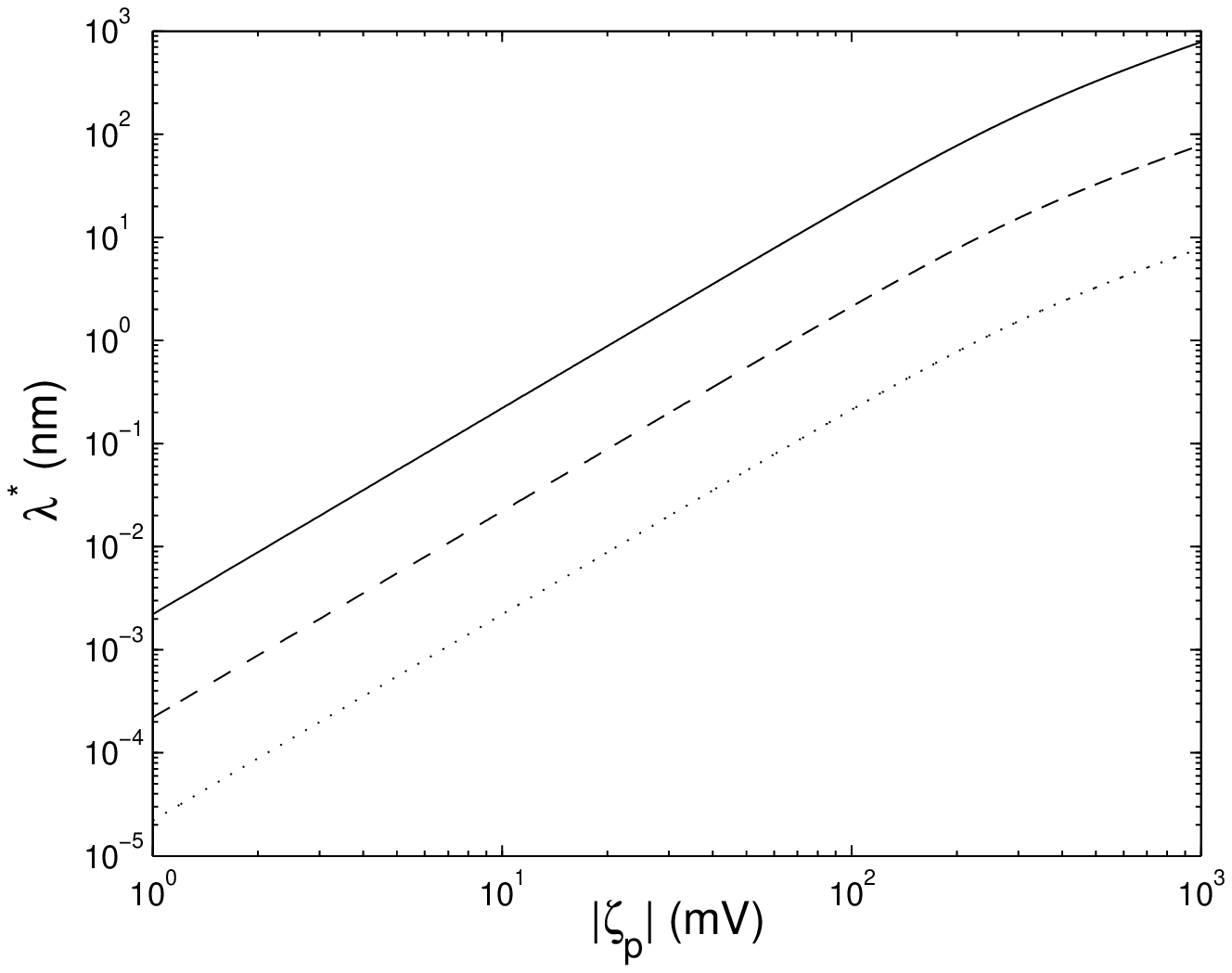}
\caption{Left: variation of the apparent slip length $\lambda$
\eqref{slip1} for pure water as a function of the wall zeta
potential $\zeta_w$ for $\zeta_p=50$ mV,
$n_0=10^{-6}$~mol~l$^{-1}$ (pure water), $\kappa h=10$ and $\kappa
a \ll 1$; the slip length reaches a maximum $\lambda^*$ for
$\zeta_w=\zeta_w^*$. Right: maximum value of the apparent slip
length $\lambda^*$ as a function of the particle zeta potential
$\zeta_p$ for $\kappa h=10$, $\kappa a \ll 1$ and three values of
the ionic strength: $n_0=10^{-6}$~mol~l$^{-1}$ (pure water,
$\kappa^{-1}\approx 300$ nm, solid line),
$n_0=10^{-4}$~mol~l$^{-1}$ ($\kappa^{-1}\approx 30$ nm, dashed
line), $n_0=10^{-2}$~mol~l$^{-1}$ (tap water, $\kappa^{-1}\approx
3$ nm, dotted line).} \label{maximum}
\end{figure}

Finally, we note the apparent slip length \eqref{slip1} vanishes
for two values of the wall zeta potential: $\zeta_w=0$ and
$\zeta_w=\zeta_p/f(\kappa a)$. Consequently, in between these two
values, the slip length reaches a maximum value $\l$ when the wall
zeta potential is equal to $\zeta_w=\zeta_m^*$, {\it i.e.} $\d
\lambda / \d\zeta_w(\zeta_m^*)=0$. This is illustrated in Figure
\ref{maximum} (left).

\subsection{Order of magnitude for water}

Let us address here the case of water at room temperature
(T=300\deg C, $\epsilon$=80, $\ell \approx 2$~\AA). We have
calculated numerically the maximum apparent slip lengths which
could be obtained in an experiment, $\lambda^*$, as a function of
the particle zeta potential $\zeta_p$. The results are displayed
in Figure \ref{maximum} (right). We first note that $\lambda^*$
increases with $|\zeta_p|$. Furthermore, the maximum slip length
can take values as low as molecular sizes or below and, in the
case of pure water, can be as high as hundreds of nanometers.

The data for the low values of $|\zeta_p|$ display a power-law
behavior, which we can analyze as follows. Let us consider formula
\eqref{slip1}. The two terms in the denominator will be of the
same order of magnitude if $t_w$ is larger than a critical value
$\tilde{t}_w$ which is given by
\begin{equation}\label{critical}
\tilde{t}_w\approx \left(\frac{1}{1+ \frac{48\pi \ell \e k_B
T}{e^2 \kappa h}}\right)^{1/2},
\end{equation}
where we have used \eqref{cond2} to relate the conductivity to the
screening length. The smallest value of \eqref{critical} will be
obtained, say, for $\kappa h \approx 10$, in which case we get
$\tilde{t}_w \approx 0.86$ which corresponds to a critical wall
zeta potential $\tilde{\zeta}_w\approx 135$ mV. Consequently, when
$\zeta_w\lesssim \tilde{\zeta}_w$, \eqref{slip1} can be simplified
to
\begin{equation}\label{}
\frac{\lambda}{h}=\frac{\zeta_w(f(\kappa
a)\zeta_p-\zeta_w)(\e)^2}{\sigma\mu h^2},
\end{equation}
for which it is easy to get
\begin{equation}\label{exp}
\zeta_w^*=\frac{f(\kappa a)}{2}\zeta_p \,,\quad
{\lambda^*}=\frac{(\e f(\kappa a)\zeta_p)^2}{4\sigma\mu h}\cdot
\end{equation}
The exponent 2 given by equation \eqref{exp} agrees well with the
power-law data presented in Figure \ref{maximum} (right).

\subsection{Comparison with experiments}

Two comparisons with experimental results can now be given. First,
we wish to comment on the general order of magnitude of the slip
lengths obtained. For a review of the pressure-driven flow
experiments in capillaries which report some degree of slip as
summarized in the introduction, the reader is referred to
\cite{LaugaStone}.

The order of magnitude for the maximum slip lengths given by our
mechanism (tens to hundreds of nanometers) are consistent with the
slip lengths measured in the indirect pressure-driven slip
experiments of \cite{Churaev,Cheng,Breuer}. Of course, the effect
we report here does not directly apply to their pressure drop
versus flow rate measurements, but the comparison shows that both
effects are comparable in magnitude and therefore the apparent
slip mechanism could have important consequences on experimental
probing of the no-slip boundary condition.

We also wish to address specifically the experiment of Tretheway
\& Meinhart \cite{Meinhart} for which our study directly applies.
The channels used in their experiment have height $2h=30~\mu$m and
width $2w=300~\mu$m; the separation of scale $w\gg h$ allows us to
approximate the flow by that between two parallel plates with $h=
15$~$\mu$m. Details of the electrical characteristics of the water
used in the experiment were not reported, but the water was
deionized; we will therefore assume that the ion concentration was
small and will take it to be that of pure water $n_0\approx
10^{-6}$~mol~l$^{-1}$ for which $\kappa^{-1}\approx 300$ nm, so
that $\kappa h \approx 50$. Particles with radius $a=$150 nm were
used in the P.I.V. system, so that $\kappa a \approx 1/2$, for
which we will approximate $f(\kappa a)\approx 2/3$. If we assume
$|\zeta_p|=10$ mV, we obtain that $\lambda^*$ is essentially zero.
If however $|\zeta_p|=50$ mV, we get $\lambda^*\approx 1$ nm and
$|\zeta_p|=200$ mV leads to $\lambda^*\approx 18$ nm. Although
beyond molecular size, these values are much too small to explain
the data reported in \cite{Meinhart} where $\lambda \approx
1$~$\mu$m. As a consequence, we can conclude that the effect
reported here is probably not responsible for the large slip
length observed in \cite{Meinhart}. Alternative mechanisms would
have to be invoked to explain the data, such as the presence of
surface attached bubbles \cite{LaugaStone}.

\section{Conclusion}

We have reported in this paper the following new mechanism. When
small charged colloidal particles are used in a pressure-driven
flow experiment to probe the profile of the velocity field of an
electrolyte solution (e.g. P.I.V. in water), their velocities may
include an ``apparent slip'' component even though the velocity
field in the fluid does not violate the no-slip boundary
condition. This apparent slip is in fact an electrophoretic
velocity for the particles which are subject to the streaming
potential, {\it i.e.}, the flow-induced potential difference that
builds up along the channel due to the advection of free screening
charges by the flow. A similar effect is expected to occur in
shear-driven flows.

The expected maximum orders of magnitude for the apparent slip
lengths were given under normal conditions in water. Although the
effect was found to be too small to explain the data reported in
\cite{Meinhart}, its magnitude is consistent with other indirect
investigations of fluid slip in pressure-driven flow experiments.
As a consequence, the analysis presented here could be a useful
tool for experimentalists by allowing them to estimate
quantitatively the importance of this apparent slip in their
experiments.

The idea that free passive particles could go faster than the
surrounding flowing liquid, although counter-intuitive at first,
is in fact not unnatural: a similar phenomenon occurs in
electrophoresis where, beyond the double layer, the ambient liquid
is at rest. We also note from equation \eqref{slip1} and the
scalings presented above that the effect increases when the ionic
strength of the solution, and therefore its conductivity,
decreases; this is because flow of an electrolyte with low ion
concentration will necessary lead to the induction of a large
streaming electric field to counteract the advection-of-charge
electric current.

The model chosen for the calculations used several simplifying
assumptions. Our calculations were two-dimensional and we
neglected in the model the effect of surface conductance as well
as interactions between particles. We also assumed that the
streaming electric field was uniform on the length scale of the
particle and its double layer. We do not expect that relaxing
these assumptions would change qualitatively the physical picture
introduced in this paper.

\section*{Acknowledgments}
We thank Shelley Anna, Michael Brenner, Henry Chen, Todd Squires,
Howard Stone, and Abraham Stroock for useful suggestions and
stimulating discussions. Funding by the Harvard MRSEC is
acknowledged.

\end{document}